\documentclass[12pt]{article}
\usepackage{epsfig}
\usepackage{comment}
\usepackage{latexsym}
\usepackage{color}
\usepackage{amsmath}
\usepackage{hyperref}

\newcommand{\mysquare}[0]{\raise-.2ex\hbox{{\Large$\Box$}}}
\def\lsim{\mathrel{\rlap {\raise.5ex\hbox{$ < $}}
{\lower.5ex\hbox{$\sim$}}}}
\def\gsim{\mathrel{\rlap {\raise.5ex\hbox{$ > $}}
{\lower.5ex\hbox{$\sim$}}}} \topmargin -1.5cm \textheight=22.5cm \textwidth=16.5cm
\catcode`\@=11
\newcount\hour
\newcount\minute
\newtoks\amorpm
\hour=\time\divide\hour by60 \minute=\time{\multiply\hour by60 \global\advance\minute by-\hour}
\edef\standardtime{{\ifnum\hour<12 \global\amorpm={am}%
        \else\global\amorpm={pm}\advance\hour by-12 \fi
        \ifnum\hour=0 \hour=12 \fi
        \number\hour:\ifnum\minute<10 0\fi\number\minute\the\amorpm}}
\edef\militarytime{\number\hour:\ifnum\minute<10 0\fi\number\minute}
\def\draftlabel#1{{\@bsphack\if@filesw {\let\thepage\relax
   \xdef\@gtempa{\write\@auxout{\string
      \newlabel{#1}{{\@currentlabel}{\thepage}}}}}\@gtempa
   \if@nobreak \ifvmode\nobreak\fi\fi\fi\@esphack}
        \gdef\@eqnlabel{#1}}
\def\@eqnlabel{}
\def\@vacuum{}
\def\draftmarginnote#1{\marginpar{\raggedright\scriptsize\tt#1}}
\def\draft{\oddsidemargin -.2truein
        \def\@oddfoot{\sl preliminary draft \hfil
        \rm\thepage\hfil\sl\today\quad\militarytime}
        \let\@evenfoot\@oddfoot \overfullrule 3pt
        \let\label=\draftlabel
        \let\marginnote=\draftmarginnote
   \def\@eqnnum{(\theequation)\rlap{\k

 ern\marginparsep\tt\@eqnlabel}%
\global\let\@eqnlabel\@vacuum}  }

\newcommand{\be}[0]{\begin{equation}}
\newcommand{\ee}[0]{\end{equation}}
\newcommand{\ba}[0]{\begin{eqnarray}}
\newcommand{\ea}[0]{\end{eqnarray}}

\def\bs{\begin{subequations}}
\def\es{\end{subequations}}

\def\thebibliography#1{%
\vskip 0.5cm \centerline{\bf \Large References}
\list{%
[\arabic{enumi}]}{\settowidth\labelwidth{[#1]} \leftmargin\labelwidth \advance\leftmargin\labelsep
\usecounter{enumi}}
\def\newblock{\hskip .11em plus .33em minus .07em}
\sloppy\clubpenalty4000\widowpenalty4000 \sfcode`\.=1000\relax}

\renewcommand{\section}{\setcounter{equation}{0}\@startsection
{section}{1}{0mm}{-\baselineskip}{0.5\baselineskip} {\normalfont\Large\bfseries}}

\renewcommand{\subsection}{\@startsection
{subsection}{2}{0mm}{-\baselineskip}{0.5\baselineskip} {\normalfont\large\bfseries}}

\renewcommand{\subsubsection}{\@startsection
{subsubsection}{3}{0mm}{-\baselineskip}{0.5\baselineskip} {\normalfont\normalsize\slshape}}

\usepackage{amssymb,amsfonts}
\usepackage{graphicx}
\usepackage{cite}

\newcommand{\Z}{\mathbb{Z}}

\renewcommand{\O}{{\cal O}}

\newcommand{\abs}{|}
\renewcommand{\o}{\overset{\circ}}
\newcommand{\oo}{\overset{\circ\circ}}

\newcommand{\where}{\mbox{where}}
\newcommand{\with}{\mbox{with}}

\renewcommand{\and}{\mbox{and}}

\newcommand{\N}{{\cal N}}
\newcommand{\C}{{\cal C}}
\newcommand{\A}{{\cal A}}
\newcommand{\V}{{\cal V}}

\newcommand{\dil}{{\phi_{\mbox{\tiny dil}}}}
\newcommand{\dilp}{{\phi'_{\mbox{\tiny dil}}}}

\newcommand{\tk}{\tilde k}

\newcommand{\tn}{\tilde n}

\newcommand{\bea}{\begin{eqnarray}}
\newcommand{\eea}{\end{eqnarray}}
\newcommand{\dis}{\displaystyle}

\begin{document}
\begin{titlepage}
\begin{flushright}
CPHT--PC010.0210\\
March 2010
\end{flushright}

\vspace{.6cm}

\begin{centering}
{\bf\huge Attractions to radiation-like eras in \\\vspace{2mm}superstring cosmologies}\\

\vspace{8mm}
 {\Large Herv\'e Partouche}

\vspace{4mm}

{Centre de Physique Th\'eorique, Ecole Polytechnique,$^\dag$
\\
F--91128 Palaiseau cedex, France\\
{\em Herve.Partouche@cpht.polytechnique.fr}}

 \vspace{6mm}

{\it Based on a talk given at the ``9th Hellenic School and Workshops on Elementary Particle Physics and Gravity'', Corfu, Greece, August 30 - September 20, 2009.}

\end{centering}

\vspace{2mm}
$~$\\
\centerline{\bf\Large Abstract}
\vskip .3cm
\noindent
We review the cosmology induced by finite temperature and quantum effects on non-super\-symmetric string models. We show the evolution is attracted to radiation-like solutions after the Hagedorn era and before the electroweak phase transition. This mechanism generates a hierarchy between the Planck mass and the supersymmetry breaking scale. A dynamical change of space-time dimension can take place.  


\vspace{3pt} \vfill \hrule width 6.7cm \vskip.1mm{\small \small \small
\noindent
 $^\dag$\ Unit{\'e} mixte du CNRS et de l'Ecole Polytechnique,
UMR 7644.}

\end{titlepage}
\newpage
\setcounter{footnote}{0}
\renewcommand{\thefootnote}{\arabic{footnote}}
 \setlength{\baselineskip}{.7cm} \setlength{\parskip}{.2cm}

\setcounter{section}{0}

The observations of the Cosmic Microwave Background at 3K show that a small percent of the energy density of the universe is due to a thermal bath of radiation. Because of its expansion, our space was smaller and hotter in the past and a radiation dominated era must have taken place. In such a state, the universe looks like a black body, which is commonly described as a ``big 3-dimensional box", with periodic boundary conditions, filled with a gas of massless states at temperature $T$. 

With this simple picture in mind, we describe our space as a torus $T^3$ of volume $V=(2\pi R_{\rm box})^3$, where the thermal bath of states contains in general modes of various masses. Moreover, the pressure of this gas pushes the ``walls of the box'' and induces an evolution of the universe. In some cases, the volume $V$, the temperature $T$, the pressure $P$ and the energy density $\rho$ of the bath may be computed for all time, supposing their evolutions are {\rm quasi-static}. 

Since black body physics is quantum, our aim is to implement the above approach of cosmology in a well defined framework for quantum gravity, namely string theory. At present time, it is the most promising context for describing the very early phases of the universe, with the hope to provide a consistent description of the epochs where general relativity predicts a big bang initial singularity. Moreover, string theory unifies gravity to the other known interactions. Thus, it has potential for connecting two  disciplines: Cosmology and particle physics. Actually, at late time, when the temperature is low, one would like to recover models where $\N=1$ supersymmetry in 4 dimensions is spontaneously broken. Since we are interested in this review in homogeneous and isotropic evolutions, the models we consider have at least 3 mass scales a priori time-dependent: The temperature $T(t)$, the supersymmetry breaking scale $M(t)$ and the inverse of the scale factor $1/a(t)$. 

In this context, we can schematically divide the cosmological evolution in three epochs:

{\bf - The Hagedorn era :} At ultra high temperature, a phase transition occurs generically in string theory. It is due to the exponential growth of the number of 1-particle states as a function of mass. This implies a divergence of the canonical partition function at the Hagedorn temperature $T_H$ that signals a phase transition. Several strategies to deal with it have been proposed in the literature but will not be discussed here \cite{Hage}. 

{\bf - The electroweak phase :} When the temperature approaches the renormalization group invariant electroweak scale $Q$, radiative corrections  become important. The latter are responsible for the electroweak symmetry breaking via the Higgs mechanism. It has been shown at the level of field theory  that the supersymmetry breaking scale $M(t)$ is then dynamically  stabilized around $Q$ \cite{M=Q}.

{\bf - The intermediate era :} In the present work, we focus on the cosmological period where the  temperature and supersymmetry breaking scale are in the ranges $T_H\gg T\gg Q$, $T_H\gg M\gg Q$. In practice, we consider arbitrary initial boundary conditions at the exit of the Hagedorn era and study the general isotropic and homogeneous evolutions of the universe. We are going to see that they are attracted to Radiation-like Dominated Solutions (RDS) \cite{attrac, RDS, Thfield}.


To be specific, we consider at tree level heterotic string models with $\N=2$ or $\N=1$ supersymmetry in 4 dimensions, with internal space $T^2\times T^4/\Z_2$ or $T^6/(\Z_2\times \Z_2)$, respectively. These choices have the advantage to allow an exact (in $\alpha'$) conformal field theory description on the world sheet, even when  the models are deformed in order to break spontaneously $\N=2$ or $1\to 0$. Physically, this can be achieved by switching on magnetic fluxes in the internal space, using a stringy generalization of the Scherk-Schwarz mechanism \cite{SSstring}. In practice, this is done by imposing different boundary conditions between superpartners, along internal 1-cycles. For example, we implement this around  the circle $S^1(R_4)$ of radius $R_4$ on which the orbifold may act. 

To study the canonical ensemble of this non-supersymmetric spectrum, we consider the Euclidean version of the background, where the time is compactified on a circle $S^1(R_0)$ of perimeter $\beta=2\pi R_0$. To interpret $\beta$ as the inverse temperature (measured in string frame), we impose (anti-)periodic boundary conditions for the bosons (fermions) along this circle.
The free energy $F$ defined in terms of the canonical partition function ${\cal Z}_{\rm th}$ can be expressed as $F=-{(\ln {\cal Z}_{\rm th})/ \beta}= -{Z/ \beta}$, 
where $Z$ is the string partition function computed on the Euclidean background. From now on, we suppose the string coupling is small and compute $Z$ at 1-loop order, which means the gas is almost perfect. The benefit to evaluate $Z$ in string theory  is that it is finite in the UV. In field theory, this would only be the case for supersymmetric spectra. 

In our explicit heterotic backgrounds, one finds
\be
\label{p}
{Z\over \beta V}={1\over \beta^4}\,  p(z)\quad \where\quad p(z)=n_T\, f_T(z)+ n_V\, f_V(z)+ \tn_T\, c_4+\cdots\;,  \quad e^z\equiv {R_0\over R_4}\; , 
\ee
and $c_D={\Gamma(D/2) \pi^{-D/2}}\sum_{\tk_0}\abs 2\tk_0+1\abs^{-D}$. The integer $n_T+\tn_T$ is the number of massless boson/fermions pairs in the parent supersymmetric model we started with. The important thing is that $\tn_T=0$ when the orbifold action is trivial on the direction 4, while $\tn_T>0$ when it is non-trivial. $n_V$ is an integer that depends on the details of the internal flux and satisfies $-n_T<n_V\le n_T$. The contributions explicitly given in Eq. (\ref{p}) arise from the Kaluza Klein (KK) towers of states associated to the radii $R_0$ and $R_4$, which by definition are large compared to 1 (in string length unit) in the intermediate era. The expressions of $f_T$ and $f_V$ can be found in  \cite{attrac, RDS}. The remaining contributions in Eq. (\ref{p}) arise from the other massive states. Supposing all internal space moduli (except $M$) introduce mass scales much larger than $T$ and $M$, these terms are found to be of order $e^{-2\pi R_0}$ or $e^{-2\pi R_4}$ and can be safely neglected. This fact is just the consequence of the Boltzmann factor suppression. 

To study the back-reaction of the 1-loop correction at finite temperature on the classically flat and static background, we consider the low energy effective action. In the intermediate era, the latter is well defined in four dimensions, since the universe is already very large (compared to the string length). For the Lorentzian metric in Einstein frame, dilaton $\dil$ and scalar $R_4$, the action takes the explicit form,
\be
\label{S}
S=\int d^4x \sqrt{-g}\left( {R\over 2}-{1\over 2}(\partial \Phi)^2-{1\over 2} (\partial \phi_\bot)^2+e^{4\dil}{Z\over \beta V}\right)\, ,
\ee
where $\Phi= \sqrt{2/ 3} (\dil-\ln R_4)$ and $\phi_\bot={1\over \sqrt{3}}(2\dil+\ln R_4)$. The temperature and supersymmetry breaking scale measured in Einstein frame are $T=1/(2\pi R_0 e^{-\dil})$ and $M=e^{\sqrt{3/2}\Phi}=1/(2\pi R_4e^{-\dil})$. For an homogeneous isotropic universe, the metric takes the form $ds^2=-T(x^0)^{-2}(dx^0)^2+a(x^0)^2[(dx^1)^2+(dx^2)^2+(dx^3)^2]$, where $a=2\pi R_{\rm box}e^{-\dil}$. Note that the laps function is the inverse temperature, as follows by analytic continuation of the Euclidean background. The stress-energy tensor contains the classical scalar kinetic terms and a 1-loop contribution ${T_\mu}^\nu={\rm diag}{(-\rho,P,P,P)_\mu}^\nu$ that satisfies
\be
P=-\left({\partial F\over \partial V}\right)_\beta e^{4\dil}\; , \qquad \rho={1\over V}\left({\partial (\beta F)\over \partial \beta}\right)_V e^{4\dil}\, .
\ee
It is remarkable that these relations derived from the variational principle coincide with the results found from the laws of thermodynamics. In the present case, we have $P=T^4p(z)$ and $\rho=T^4r(z)$, where $r=3p-dp/dz$ and $e^z=M/T$. 

The equations of motion yield
\be
\label{eqs}
 T={\A(z)\over a}\; ,\quad \qquad  H={\C_1\over a^4}\; ,\quad \qquad \left\{
 \begin{array}{l}
 \C_2 \oo z+ \C_3 \o z{}^2+\C_4 \o z+\dis {d\V\over d z}=0\\ 
 \C_5 \oo \phi_\bot+ \C_6 \o \phi_\bot=0
 \end{array}
 \right. ,
 \ee
where $H=\dot a/a$ with the definition of time $dt=dx^0/T$. We also denote $\dis\dot y\equiv H({dy/d\ln a}):=H\o y$. In Eqs (\ref{eqs}), the functions $\C_i(z; \o z,\o \phi_\bot)$ $(i=1,\dots, 6)$ can be found in \cite{attrac} and  ${d\V/d z}=r-4p$. Clearly, when $\V(z)$ admits an extremum $z_c$, a very simple solution with constant $z\equiv z_c$ and $\phi_\bot\equiv \phi_{\bot 0}$ exists if $\A(z_c)>0$,
\be
\label{rad}
M(t)=T(t)\times e^{z_c}={1\over a(t)}\times e^{\A(z_c)}\quad \with \quad a(t)=\sqrt{t}\times \big(4\A(z_c)\big)^{1/4} ,\quad \phi_\bot(t)=\phi_{\bot 0}\, .
\ee
Geometrically, it satisfies $R_0(t)\propto R_{\rm box}(t)\propto R_4(t)\propto e^{-2\dil (t)}\to +\infty$, consistently with the  weak coupling approximation. Moreover, all time-derivatives vanish at late time, so that the quasi static hypothesis is also reliable. Note that the evolution (\ref{rad}) mimics that of a radiation dominated universe in 4 dimensions, but {\em is not radiation dominated}. Actually, since $R_4$ increases proportionally to $R_0$, the masses of the KK states along the direction 4 decrease with $T$. Thus,  these modes never decouple from the thermal bath. In fact, the state equation is $\rho=4P$ and it is only once we take into account the classical kinetic energy of $M(t)$ that the {\em total} energy density and pressure satisfy Stefan's law in 4 dimensions, $\rho_{\rm tot}=3P_{\rm tot}$.

When $n_V>0$, one finds that $\V(z)$ is monotonically decreasing and unbounded from below. As a consequence, there is no critical point $z_c$ and the above radiation-like solution (RDS) does not exist. In this case, the universe is actually attracted into a phase of contraction where our analysis breaks down \cite{attrac}. Thus, we focus from now on the models where $n_V<0$. When the orbifold action is non-trivial on the direction 4 (and thus $\tn_T>0$), one finds $\V(z)$ has a unique extremum $z_c$, which is a minimum such that $\A(z_c)>0$. As a result, the RDS in Eq. (\ref{rad}) exists. However, the same conclusions apply to the models where the orbifold action is trivial on $S^1(R_4)$ (and thus $\tn_T=0$), if and only if $-1/15<n_V/n_T<0$. 

To be relevant, the RDS we have found have to be stable under small time-dependent perturbations, $z=z_c+\varepsilon_{(z)}$, $\phi_\bot=\phi_{\bot0}+\varepsilon_{(\bot)}$. This fact happens to be true since the equations of motion imply $\varepsilon_{(z)}$ and $\varepsilon_{(\bot)}$ are dying off, when $t\to +\infty$. The important thing to note is that in some sense, the RDS is actually the unique cosmological evolution in the intermediate era. 
For arbitrary initial conditions at the exit of the Hagedorn era, the generic solution of the system of differential equations converges towards the RDS (as can be observed by numerical simulation). The latter is thus a global attractor of the dynamics. 

When the orbifold action is trivial on $S^1(R_4)$ and $n_V/n_T\le -1/15$, $\V(z)$ is  monotonically increasing. It diverges exponentially when $z\to +\infty$ and  converges to a constant when $z\to -\infty$. This suggests that the dynamics may attract $z(t)$  to large negative values, where it should behave as a modulus along the almost flat direction of $\V$.  Actually, for $e^z\equiv R_0/R_4\ll 1$, one finds
\be
\label{z<-1}
{Z\over \beta V (2\pi R_4)}={1\over \beta^5}\left(n_T\,  c_5+\left({R_0\over R_4}\right)^4\left({n_T\over 15}+n_V\right){c_4\over 2}+\cdots\right) ,
\ee 
where terms of order $e^{-2\pi (R_0/R_4)^{-1}}$ are neglected. The above expression invites us to reconsider the dynamics of the universe from a 5 dimensional point of view, where $S^1(R^4)$ is interpreted as a fifth ``external'' dimension. Thus, we rewrite the action (\ref{S}) in terms of fields suitably normalized in 5 dimensions and denoted with "primes",
\be
\label{S5}
S=\int d^5x \sqrt{-g'} \left({R'\over 2}-{2\over 3}(\partial \dilp)^2+e^{{10\over 3}\dilp}{Z\over \beta V (2\pi R_4)}\right)\, .
\ee
Looking for homogenous and {\em anisotropic} cosmologies, we consider a metric ansatz $ds^{\prime 2}=-T'(x^0)^{-2}(dx^0)^2+a'(x^0)^2\left[(dx^1)^2+ (dx^{2})^2+ (dx^{3})^2\right]+b(x^0)^2(dx^4)^2$, where $T'=1/(2\pi R_0 e^{-2\dil / 3})$, $a'=2\pi R_{\rm box} e^{-2\dilp/ 3}$, $b=2\pi R_4 e^{-2\dilp/ 3}$. 
The equations of motion for the time-dependent fields $T$, $a'$, $\dilp$ and $e^\xi:= b/a'$ can be found in  \cite{attrac}. To solve them in the regime $R_0/R_4\ll1$, we first neglect the subdominant term of order 4 in  Eq. (\ref{z<-1}). 
It is then possible to show that $e^z$, $\dilp$, $\xi$ are converging to constants $e^{z_0}\ll 1$, $\phi'_{\mbox{\tiny dil}0}$, $\xi_0$, when $t\to +\infty$. This parameters are actually moduli determined by the arbitrary choice of initial conditions at the exit of the Hagedorn era. The dynamics is thus attracted by the solution
\be
\label{att}
b(t)=a'(t)\times e^{\xi_0}={1\over T'(t)}\times e^{-z_0}\; , \quad \dilp=\phi'_{\mbox{\tiny dil}0} \quad \where \quad 6H^{\prime 2}={1\over a^{\prime 5}}\times 4n_Tc_5e^{-5(z_0+\xi_0)}\, ,
\ee
which describes a radiation dominated universe in 5 dimensions. This is not a surprise, since the scalar kinetic energies vanish and the state equation of the total energy density and pressure reduces to the thermal one, $\rho'=4P'=4 T^{\prime 5} n_T c_5$, which satisfies Stefan's law in 5 dimensions. 

To study the effect of the residual term $\O(e^{4z})$ we have neglected in Eq. (\ref{z<-1}), we consider small perturbations around the evolution (\ref{att}), 
$z=z_0+\varepsilon_{(z)}$, $\xi=\xi_0+\varepsilon_{(\xi)}$, $\dilp=\phi'_{\mbox{\tiny dil}0} +\varepsilon_{\mbox{\tiny dil}}$. Using the equations of motions at first order in $e^{4z}$, one finds $\o\varepsilon_{(z)}=e^{4z_0}{3c_4\over 2c_5}\left({n_V\over  n_T}+{1\over15}\right)<0$, $\o\varepsilon_{(\xi)}=-{4\over 3}\o\varepsilon_{(z)}$, $\o\varepsilon_{\mbox{\tiny dil}}=0$.
This shows that  the residual force implies $e^z$ decreases slightly, a fact that justifies the validity of the attraction mechanism towards the RDS in Eq. (\ref{att}). Actually, the process we have described is a dynamical decompactification of the internal $S^1(R_4)$, with a ``dilution'' of the supersymmetry breaking flux around it.

To summarize, we have shown that for $n_V<0$, the evolution of the universe converges to a radiation-like cosmology during the intermediate era, with an eventual change of space-time dimension. Actually, the condition $n_V<0$ means that the internal flux implies a {\em negative} contribution to $Z$. It is also important to note that even if the supersymmetry breaking scale $M(t)$ is  initially close to the Planck mass, its dynamics drives it to the electroweak scale, thus creating a hierarchy. Moreover, let us recall we derived these results under the hypothesis that all internal space moduli (except $M$) introduce mass scales much larger than $T$ and $M$. However, this fact happens to be inevitable, as a consequence of over attraction properties of the radiation-like solutions   \cite{attrac}. This fact is relevant to generate a ``desert'' between the Planck and the electroweak scales. 

\section*{Acknowledgements}

The works reviewed here are based on collaborations with F. Bourliot, T. Catelin-Jullien, J. Estes, C. Kounnas and N. Toumbas I would like to thank. I am also grateful to the organizers of this conference for the opportunity to present these results. 
This work is partially supported by the contracts PITN-GA-2009-237920,  ERC-AdG-226371, ANR 05-BLAN-NT09-573739,  CEFIPRA/IFCPAR 4104-2, PICS 3747 and 4172.


\vspace{.3cm}
\begin{thebibliography}{99}

\bibitem{Hage}
  C.~Angelantonj, C.~Kounnas, H.~Partouche and N.~Toumbas,
 ``Resolution of Hagedorn singularity in superstrings with gravito-magnetic
  fluxes,''
  Nucl.\ Phys.\  B {\bf 809} (2009) 291
  [arXiv:0808.1357 [hep-th]];

  J.~Atick and E.~Witten,
 ``The Hagedorn transition and the number of degrees of freedom of string
 theory,''
  Nucl.\ Phys.\  B {\bf 310}, 291 (1988);

  I.~Antoniadis, J.~P.~Derendinger and C.~Kounnas,
  ``Non-perturbative temperature instabilities in N = 4 strings,''
  Nucl.\ Phys.\  B {\bf 551} (1999) 41
  [arXiv:hep-th/9902032].

\bibitem{M=Q}
  C.~Kounnas, F.~Zwirner and I.~Pavel,
 ``Towards a dynamical determination of parameters in the minimal
  supersymmetric standard model,''
  Phys.\ Lett.\  B {\bf 335} (1994) 403
  [arXiv:hep-ph/9406256].


\bibitem{attrac}
  J. Estes,  C.~Kounnas and H.~Partouche,
  ``Superstring cosmology for $\N_4 = 1\rightarrow 0$ superstring vacua,'' 
  arXiv:1003.0471 [Unknown];
 
  F.~Bourliot, C.~Kounnas and H.~Partouche,
  ``Attraction to a radiation-like era in early superstring cosmologies,''
  Nucl.\ Phys.\  B {\bf 816} (2009) 227
  [arXiv:0902.1892 [hep-th]];

  F.~Bourliot, J.~Estes, C.~Kounnas and H.~Partouche,
 ``Cosmological phases of the string thermal effective potential,''
  Nucl.\ Phys.\  B {\bf 830} (2010) 330
  [arXiv:0908.1881 [hep-th]];

  F.~Bourliot, J.~Estes, C.~Kounnas and H.~Partouche,
  ``Thermal and quantum induced early superstring cosmology,''
  arXiv:0910.2814 [Unknown].
  
  
  
  
\bibitem{RDS}
  T.~Catelin-Jullien, C.~Kounnas, H.~Partouche and N.~Toumbas,
  ``Thermal/quantum effects and induced superstring cosmologies,''
  Nucl.\ Phys.\  B {\bf 797} (2008) 137
  [arXiv:0710.3895 [hep-th]];
  
  T.~Catelin-Jullien, C.~Kounnas, H.~Partouche and N.~Toumbas,
 ``Induced superstring cosmologies and moduli stabilization,''
  Nucl.\ Phys.\  B {\bf 820} (2009) 290
  [arXiv:0901.0259 [hep-th]];

  T.~Catelin-Jullien, C.~Kounnas, H.~Partouche and N.~Toumbas,
 ``Thermal and quantum superstring cosmologies,''
  Fortsch.\ Phys.\  {\bf 56} (2008) 792
  [arXiv:0803.2674 [hep-th]].
  
    
\bibitem{Thfield}
  C.~Kounnas and H.~Partouche,
  ``Inflationary de Sitter solutions from superstrings,''
  Nucl.\ Phys.\  B {\bf 795} (2008) 334
  [arXiv:0706.0728 [hep-th]];
  
  C.~Kounnas and H.~Partouche,
 ``Instanton transition in thermal and moduli deformed de Sitter cosmology,''
  Nucl.\ Phys.\  B {\bf 793} (2008) 131
  [arXiv:0705.3206 [hep-th]].
  
  
  
\bibitem{SSstring}
  C.~Kounnas and M.~Porrati,
  ``Spontaneous supersymmetry breaking in string theory,''
  Nucl.\ Phys.\  B {\bf 310} (1988) 355.
  
 
  

\end{thebibliography}
\end{document}